\documentclass[prl,twocolumn,showpacs,preprintnumbers,amsmath,amssymb]{revtex4}

\usepackage{graphicx}
\usepackage{dcolumn}
\usepackage{bm}
\usepackage[]{ams,amsmath,amssymb,epsfig,color}

\begin{document}


\title{Revivals, collapses and magnetic-pulse generation in
quantum
 rings}

\author{A.S. Moskalenko}
 \email{moskalen@mpi-halle.de}
 \altaffiliation[Also at ]{A.F. Ioffe Physico-Technical Institute,
 194021 St. Petersburg, Russia}
\author{A. Matos-Abiague}%
\author{J. Berakdar}
\affiliation{%
Max-Planck Institut f\"{u}r Mikrostrukturphysik, Weinberg 2, 06120
Halle, Germany
}%

\date{\today}

\begin{abstract}
 Using a microscopic theory
 based on the density matrix formalism
we investigate  quantum revivals and collapses of  the charge
polarization and charge current dynamics in mesoscopic rings
driven by short asymmetric electromagnetic pulses. The collapsed
state is utilized for sub-picosecond switching of the current and
associated magnetization, enabling thus the generation of pulsed
 magnetic fields with a tunable time structure and
shape asymmetry which provides a new tool to study  ultrafast
spin-dynamics and ratchet-based effects.
\end{abstract}

\maketitle


\emph{Introduction.-} The spectacular advances in the design and
tunability of the time-structure, amplitude, phase, and the shape
of electromagnetic pulses
\cite{4apl2,tielking,bensky,bucksbaum,krausz,brixner} have opened
new avenues for their utilization in fundamental and applied
research. Inducing and monitoring non-equilibrium states  as they
build up and decay \cite{huber,haug_koch} are just but one example
of recent applications. This progress in laser physics is
paralleled with equally impressive development in the fabrication
and manipulation of meso and nano-scale electronic systems in
general, and of those with ring-confining geometry in particular
\cite{mailly,0n,6n,rabaud,Mohanty,Ariwala,klaui}. Subjecting
quantum rings to strongly asymmetric pulses results in the
formation of time-dependent charge polarization and charge
currents \cite{Alex_PRB2004,Alex_PRL2005}. How these
non-equilibrium phases  collapse, revive and eventually relax due
to coupling to other degrees of freedom have not yet been
addressed.
 We demonstrate that, once the evolution dynamics of the
pulse-induced charge polarization is known, a scheme can be
developed which allows to   abruptly switch on and off the charge
current in the ring, enabling thus a fine tuning of the time
structure and the shape of the light-induced current. This is
important in so far as the current in the ring is associated with
a magnetic field, i.e. a controllability of the current renders
the generation of short (down to picoseconds) magnetic pulses. The
pulse-induced magnetic field can be utilized  to study and
manipulate  locally and in a non-invasive way the picosecond
spin-dynamics \cite{Hillebrands} of magnetic samples
\cite{back,back2,siegmann,Atkinson}.
 In addition, we demonstrate how shape asymmetric
magnetic pulses can be generated and controlled revealing thus the
potential of our pulses for inducing   a ratchet effect without
spatially asymmetric potential \cite{ratchet}. The simplicity of
the proposed setup and the local tunability of the magnetic pulse
shape and duration make our scheme a valuable addition to
currently known accelerator-based methods
\cite{back,back2,siegmann} and those  based on the generation of
currents  in two photo switches excited by femtosecond laser
pulses \cite{Gerrits}.

For inducing non-equilibrium states so-called electromagnetic
half-cycle pulses (HCPs) \cite{bensky,tielking,bucksbaum} are
employed. An HCP is a highly asymmetric monocycle  pulse whose
electric field consists of a strong and short (duration
$\tau_d\gtrsim$ ps) half cycle followed by a much weaker but
longer opposite polarity half cycle. Collapses and revivals of the
HCPs-induced non-equilibrium charge polarization, the charge
current generation and suppression, and  relaxation effects are
studied using the density-matrix formalism
\cite{haug_koch,Rossi_Kuhn}. As we are interested in
non-destructive processes, weak HCPs  (few V/cm) are applied.
Thus, for low temperatures, $T$, the evolution dynamics is
governed by coupling to longitudinal-acoustic (LA) phonons.
Electron-electron interaction effects are suppressed by Pauli
blocking and energy-conservation restrictions \cite{chakraborty}.
On the other hand, as we are interested in  weak driving fields
 optical phonons do not influence the evolution
dynamics because their energy is well above  our highest excited
electron energy.

\emph{Theory.-} We consider $N$ electrons in an isolated  ring
with a mean radius $r_0$, height $z_0$, and width $d$ ($z_0 \ll
d\ll r_0$) at low  $T$. We specifically address the case where $d$
and $z_0$ are significantly smaller than the Fermi wave length of
the charge carriers. Hence,  only the
  lowest ground-state radial subbands are  populated, a situation which
  is experimentally feasible for semiconductor-based rings
\footnote{Generalization  to  multi radial channels \cite{Alex_Europhysics2005}
can be handled in a similar way.}.
The single-particle energies associated with the angular motion
(characterized by the quantum number $m$) read
 $\varepsilon_{m}
    =\hbar^{2}m^{2}/(2m^{\ast}r_{0}^{2})$,
where $m^*$ is the effective mass. The particles' angular dynamics
 is governed by the  Hamiltonian
$
  \hat{H}_{\rm tot} = \hat{H}^{\rm carr}_0 +\hat{H}^{\rm phon}_0
  +\hat{H}_{\rm P}+\hat{V}$. Here
 $\hat{H}^{\rm carr}_0=\sum_m
        \varepsilon_m
        a_m^{\dagger} a_m$
 is the free carrier Hamiltonian and $a_m^\dagger$ ($a_m$) denote
the creation  (annihilation) operators.
  $\hat{H}_0^{\rm phon}
  = \sum_{\vec{q}} \hbar \omega_{\vec{q}}
  \left( b^{\dagger}_{\vec{q}} b_{\vec{q}}
  + \frac{1}{2} \right)$ is the free-phonon Hamiltonian,
 $\omega_{\vec{q}}\;$ is the frequency of a phonon with a
momentum $\vec{q}$, and $b_{\vec{q}}^\dagger$ ($b_{\vec{q}}$) are
phonon creation (annihilation) operators. The carrier-phonon
coupling is dictated by the  Hamiltonian
$\hat{H}_P=\sum_{\vec{q},m,m'}g_{\vec{q}}^{m^\prime}b_{\vec{q}}
    a_m^\dagger a_{m-m'}+{\rm h.c.}$, where
$g_{\vec{q}}^{m-m'}=g_{\vec{q}}^{\rm bulk}\int\!\!\mbox{d}^3r
    \
    \psi_{m}^*(\vec{r})\mbox{e}^{i\vec{q}\vec{r}}\psi_{m'}(\vec{r})$
    is the electron-phonon coupling constant  \cite{Rossi_Kuhn}
    and  $\psi_{m}(\vec{r})$ are wave functions of the ring carriers in the lowest
    radial subband. The  bulk electron-LA-phonon
coupling constant $g_{\vec{q}}^{\rm bulk}$ is determined by
$|g^{\rm bulk}_{\vec{q}}|^2=\hbar|D|^2q/(2V c_{_{\rm LA}}\rho_s)$,
where $D$ is the deformation constant, $c_{\rm LA}$ is the
LA-velocity of sound, and $\rho_s$ is the lattice density. The
interaction with a time-dependent linearly polarized external
electric field $F(t)$ is given by $\hat{V}=-eF(t) r_0
\sum_{m,m'}\langle m|\cos\phi| m'\rangle a_m^\dagger a_{m'}$.
 $\phi$ identifies the electron angular position
 with respect to the pulse polarization axis  and
$e$ is the electron charge.

A key aspect of this work is that the system is excited with a
pulse which is very short on the relevant relaxation time scale.
Hence, the excitation process can be separated from relaxation and
dephasing  which occur then in a field-free manner and can be
monitored by measuring the emission spectrum.
The excitation dynamics of the  system  is governed by
Heisenberg's equation of motion (EOM) for the density operator
$\hat{\rho}_{m,m'} = a^\dagger_m a_{m'}$, i.e.,
 $
   i\hbar\partial_t\hat{\rho}_{m_1,m_2}=
   [\hat{H}^{\rm carr}_0+\hat{V},\hat{\rho}_{m_1,m_2}].
$
 In the interaction representation ($\hat{\rho}^{\rm
int}_{m_1,m_2}=\hat{\rho}_{m_1,m_2} {\rm
e}^{\frac{i}{\hbar}(\varepsilon_{m_2}-\varepsilon_{m_1})t}$) we
find
\begin{equation}\label{Eq:HOM_am1am2_simplified_interaction}
  \begin{split}
    i\hbar\partial_t\hat{\rho}^{\rm int}_{m_1,m_2}\!=&
    -\frac{1}{2}eF(t)r_0\Big(\hat{\rho}^{\rm int}_{m_1,m_2+1}
    {\rm
    e}^{\frac{i}{\hbar}(\varepsilon_{m_2}-\varepsilon_{m_2+1})t}\\
    &+\hat{\rho}^{\rm int}_{m_1,m_2-1}
    {\rm e}^{\frac{i}{\hbar}(\varepsilon_{m_2}-\varepsilon_{m_2-1})t}\\
    &-\hat{\rho}^{\rm int}_{m_1+1,m_2}
    {\rm e}^{\frac{i}{\hbar}(\varepsilon_{m_1+1}-\varepsilon_{m_1})t}\\
    &-\hat{\rho}^{\rm int}_{m_1-1,m_2}
    {\rm e}^{\frac{i}{\hbar}(\varepsilon_{m_1-1}-\varepsilon_{m_1})t}\Big)\;.
  \end{split}
\end{equation}
If the applied pulse duration   $\tau_d$ is much shorter than
$\hbar/(\varepsilon_{m+1}-\varepsilon_m)$, where $\varepsilon_m$
is in the vicinity of the  Fermi level ($E_F$) within a range
determined by $T$, then the exponential factors in
Eq.~\eqref{Eq:HOM_am1am2_simplified_interaction} hardly vary   on
the time scale $\tau_d$. For low $T$ this situation is equivalent
to $\tau_d\ll \tau_{_F}$  where $\tau_{_F}=2\pi r_0/v_{_F}$ and
$v_{_F}$ is the carrier Fermi velocity. This case (called
impulsive approximation (IA) \cite{Henriksen1999,Daems2004}) is
indeed realizable experimentally: for typical ballistic rings
$\tau_{_F}$ can be tens of picoseconds, while pulses down to
subpicosecond time scale can be generated with contemporary
techniques \cite{tielking}. Under IA,
 Eq.~\eqref{Eq:HOM_am1am2_simplified_interaction} reduces
in the basis-free form to $i\hbar\partial_t \hat{\rho}^{\rm int}=
    [-eF(t)r_0\cos\phi, \hat{\rho}^{\rm int}]$. If the
    duration of the negative HCP tail is much longer than $\tau_{_F}$
    the action of this tail on the system is averaged to zero.
Applying the pulse  at $t=0$ we operate again in the
Schr\"{o}dinger picture and derive  for the density matrix the
relation  (for $t>\tau_d$)
\begin{equation}\label{Eq:rho_basis_free_result}
    \rho_{m_1,m_2}\!=
    {\rm e}^{\frac{i}{\hbar}(\varepsilon_{m_1}-\varepsilon_{m_2})t}
    \sum_{mm'}C_{m_2,m}^* C_{m_1,m'}\rho_{m'\!,m}^0,
\end{equation}
where $\rho_{m'\!,m}^0$ is the density matrix before the application of the
pulse. $C_{m_1,m}\equiv\langle m|{\rm e}^{i\alpha\cos\phi}|m_1
\rangle=i^{m_1-m}J_{m_1-m}(\alpha)$, $J_l(x)$ are Bessel functions, and
$\alpha=r_{0} p/\hbar$, where $p=-e\int_{0}^{\tau_d} F(t) dt$ is the momentum
transferred to the charge carriers. If for $t<0$ the system is in an
equilibrium  state with an initial density matrix
$\rho_{m'\!,m}^0=f_{m}^0(\eta,T)\delta_{m'\!,m}$, where $f_{m}^0(\eta,T)$ is
the Fermi-Dirac distribution function ($\eta$ and $T$
denote the chemical potential and the temperature,
respectively) \footnote{Many-body effects on the ground-state distribution
function are suppressed due to Pauli's principle.}. For our isolated rings the
chemical potential at a given temperature is uniquely determined by the number
of particles in the ring $N$.
The pulse electric field is chosen as
 $
   F(t)=F_0\frac{t}{\tau_0}\left[\exp\left(-\frac{t^2}{2\tau_0^2}\right)
   -\frac{1}{b^2}\exp\left(-\frac{t}{b\tau_0}\right)\right].
$
The parameters $F_0$, $\tau_0$, and $b$ determine respectively the
amplitude, the duration, and the asymmetry of the pulse. The
duration $\tau_d$ of the positive half cycle is determined by
$\tau_d=\tau_0(1+\sqrt{1+2b^2\ln b^2})/b$. For $b=8$  the pulse
has basically  the experimentally observed shape and ratio of 13:1
between the maximum field values of the positive and negative
polarity parts \cite{tielking}. The peak field strength is $F_{\rm
p}\approx0.593F_0$. When rings are
  irradiated with these pulses, a charge polarization  builds up
  which is characterized by
the dipole moment $\vec{\mu}(t)=\textrm{Tr}[e \vec{r}
\hat{\rho}(t)]$. The components of the dipole moment along and
perpendicular to the pulse polarization are $\mu_\|(t)=e r_0
\sum_m \mbox{Re}[\rho_{m+1,m}]$ and $\mu_\bot(t)=e r_0 \sum_m
\mbox{Im}[\rho_{m+1,m}]$, respectively.
A detailed comparison between the dynamics of the dipole moment
calculated using the IA and the exact numerical solution for
different  $\tau_d$ but fixed  $\alpha$ endorsed our expectation
that  IA is well justified if $\tau_d$ is smaller than a quarter
period of the dipole moment oscillations (a period of oscillation
is determined by the energy difference between levels near $E_F$)
and the HCP tail duration is  longer than a quarter period of
oscillations. Thanks to the IA  a single parameter, the kick strength
$\alpha$, is sufficient for characterizing the coupling of the HCP
to the electronic system.

The non-equilibrium dynamics of the induced dipole moment is
inferred from  EOM for $\rho_{m_1,m_2}$ (including scattering from
phonons), i.e.,  \cite{Rossi_Kuhn}
\begin{equation*}\label{Eq:f_el_phon_scatt}
  \begin{split}
    i\hbar\partial_t
     \rho_{m_1,m_2}\!\!=&(\varepsilon_{m_2}-\varepsilon_{m_1})\rho_{m_1,m_2}\\
    &\!\!\!+
    \!\sum_{m_3 \vec{q}}\left[g_{\vec{q}}^{m_2-m_3}s_{\vec{q}}^{m_1,m_3}
    \!+(g_{\vec{q}}^{m_3-m_2})^{\!\ast}(s_{\vec{q}}^{m_3,m_{\!1}})^{\!\ast} \right. \\
    &\left.\hspace{0.2cm} -g_{\vec{q}}^{m_3-m_1}s_{\vec{q}}^{m_3,m_2}
    \!-\!(g_{\vec{q}}^{m_1-m_3})^{\!\ast}(s_{\vec{q}}^{m_2,m_3})^{\!\ast}\right]\!.
   \end{split}
\end{equation*}
Using the Markov approximation and neglecting polaron corrections
the phonon-assisted density matrices
$s_{\vec{q}}^{m_1,m_2}\equiv\langle b_q a^\dagger_{m_1}
a_{m_2}\rangle-\langle b_q\rangle \langle a^\dagger_{m_1}
a_{m_2}\rangle$ in  absence of an external mechanical action read
\cite{Rossi_Kuhn}
\begin{equation*}\label{Eq:dynamics_phon_assisted_density_matrixes_solution}
  \begin{split}
     s_{\vec{q}}^{m_1,m_2}\!=&-i\pi\delta(\varepsilon_{m_2}\!-\!\varepsilon_{m_1}
     \!+\!\hbar\omega_{\vec{q}})
    \sum_{m_3,m_4}(g_{\vec{q}}^{m_4-m_3})^{\!\ast}\\
    &\times\!\left[(n_{\vec{q}}\!+\!1)\rho_{m_1,m_4}\bar{\rho}_{m_3,m_2}
    -n_{\vec{q}}\rho_{m_3,m_2}\bar{\rho}_{m_1,m_4}\right],
   \end{split}
\end{equation*}
where
$\bar{\rho}_{m,m^\prime}\equiv\delta_{m,m^\prime}\!-\!\rho_{m,m^\prime}$.
For $\alpha<1$, i.e. for weak excitations,   we write
$
   \rho_{m_1,m_2} \approx
   f^0_{m_1}\delta_{m_1,m_2}+\tilde{\rho}_{m_1,m_2}.
   $
Assuming thermal equilibrium for phonons [i.e., $n_{\vec
{q}}\equiv n^0(q)$, where $n^0(q)$ is the Bose-Einstein
distribution function],  performing  the sums over $\vec{q}$,
and assuming the Debye model for the LA-phonon spectrum, we arrive
at the equation for our numerical calculations of the relaxation
dynamics
\begin{equation}\label{Eq:ac_phon_scatt_answer2}
  \begin{split}
    \partial_t
     \tilde{\rho}_{m_1,m_2}\!\!=&\left[\frac{i}{\hbar}(\varepsilon_{m_1}\!-\varepsilon_{m_2})
      -\!\sum_{m}
     \!\frac{R^{m}_{m_1}\!\!+\!R^{m}_{m_2}}{\tau_{_{\rm LA}}}\right]\tilde{\rho}_{m_1,m_2}\\
     &+\!\!\sum_{m}\!\frac{\tilde{\rho}_{m_1+m,m_2+m}}{\tau_{_{\rm
     LA}}}
     \!\left(R^{m_1}_{m_2+m}\!\!+\!R^{m_2}_{m_1+m}\right),
   \end{split}
\end{equation}
where $\tau_{_{\rm LA}}=\hbar c_{_{\rm LA}}^2\rho_s d^2
  r_0/|D|^2$,
\begin{equation}\label{Eq:R}
  \begin{split}
   R^{m}_{m'}\!\!=& {\cal F}_m^{m'}\!(q^{m}_{m'} d)\;[n^0(
     q^{m}_{m'})+f^0_{m}],\ \ \ \ q^{m}_{m'}\!
     \in\!\!\Big(0,\frac{\omega_{_{\rm D}}}{c_{_{\rm LA}}}\Big);\\
     R^{m}_{m'}\!\!=& {\cal F}_m^{m'}\!(q^{m}_{m'} d)\;[n^0(
     q^{m}_{m'})\!+\!1\!-\!f^0_{m}],\ q^{m}_{m'}\!\in\!\!\Big(\!\!-\!\frac{\omega_{_{\rm D}}}
     {c_{_{\rm LA}}},\!0\Big);\\
      R^{m}_{m'}\!\!=&0,\hspace{4.1cm} \mbox{otherwise};
  \end{split}
\end{equation}
$q^{m}_{m'}=(\varepsilon_m-\varepsilon_{m'})/(\hbar c_{_{\rm
  LA}})$  and $\omega_{_{\rm D}}$ is the Debye frequency;
  ${\cal F}_m^{m'}\!(y)=8\pi^2y\int_0^{y}\!{\rm d}x
    \frac{1}{\sqrt{1\!-\!x^2/y^2}}\;
    \frac{\sin^2\!(x/2)}{x^2\left[x^2\!-\!(2\pi)^2\right]^2}$ if
    ${\rm sgn}(m)={\rm sgn}(m')$  and ${\cal F}_m^{m'}\!(y)=0$
    otherwise.


\begin{figure}[t]
  \centering
  \includegraphics[width=7.8cm]{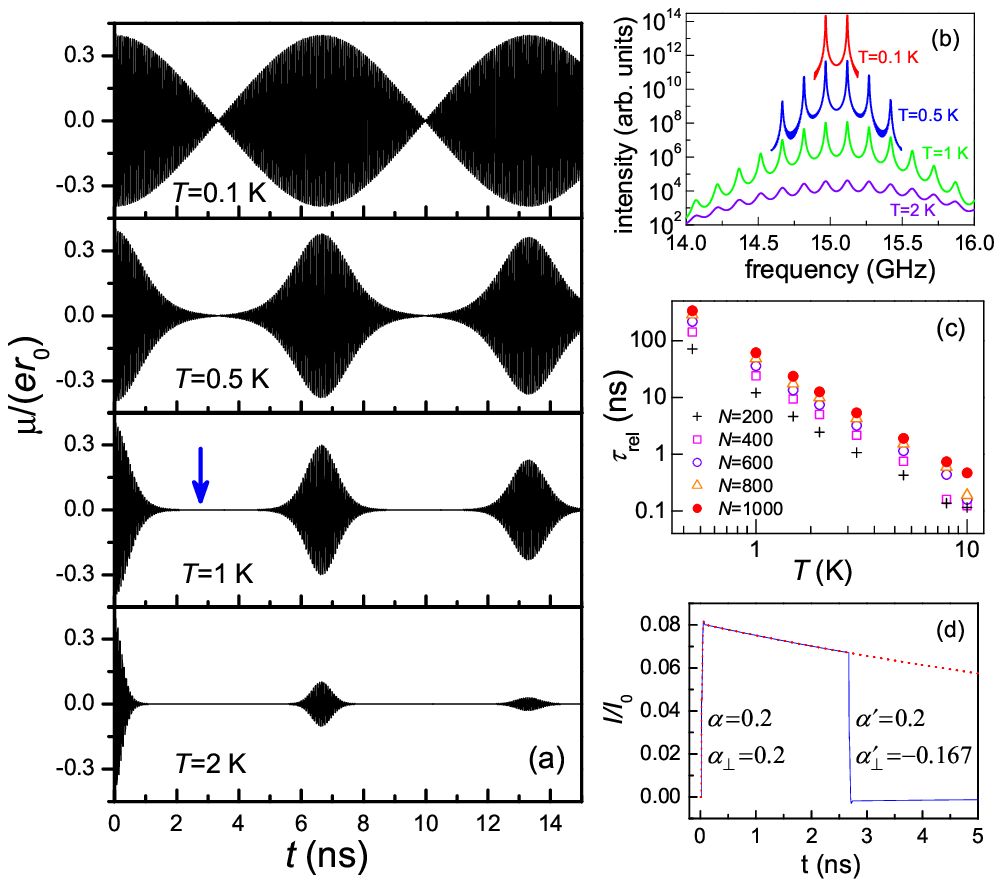}
  \caption{\label{Fig:echo_other} (a) Dynamics of {the dipole moment} $\mu$
  for different temperatures. Kick strength $\alpha=0.2$.
  (b) Emission spectra. The curves are offset vertically for
  clarity.
  (c) Dependence of
  the relaxation time of the dipole moment on temperature $T$ and the number $N$ of
  electrons.
  (d) Numerically calculated current decay (dotted line)
   after excitation by the $\frac{1}{4}T_{\rm cl}$-delayed
  sequence of two {mutually} perpendicular HCPs  and current suppression (full
  line)
  by application of the second $\frac{1}{4}T_{\rm cl}$-delayed sequence .
  Duration of each HCP is $\tau_d=3$~ps; T=1~K. The arrow in (a) marks the time moment
  when the stopping sequence of HCPs is applied.
  Parameters of the ring: $r_0=1.35~\mu{\rm m}$, $d=50~{\rm
  nm}$, and $N=400$.
  }
\end{figure}

\emph{Numerical results.-} The short-pulse induced dynamics
  occurs within the following scenario. Starting from the  ground
state a short pulse drives the system to the excited state
determined by Eq.~\eqref{Eq:rho_basis_free_result}. This
non-equilibrium state serves as the initial condition for the
numerical propagation of
equation~\eqref{Eq:ac_phon_scatt_answer2}, which in turn govern
the post-pulse charge dynamics.

For the calculations the following  ring material parameters
(corresponding to n-GaAs) are employed: $c_{\rm
LA}=4.79\times10^5~\mbox{cm/s},\ \rho_s=5.32~\mbox{g/cm}^{-3},\
|D|=8.6~\mbox{eV}, \ \hbar\omega_{_{\rm D}}=30~\mbox{meV},\
m^*=0.067m_0$ ($m_0$ is the free electron mass).
Fig.~\ref{Fig:echo_other} (a) shows the nanosecond-dynamics of the
dipole moment $\mu$  for $N=400$
  and different $T$ (the oscillatory behavior on the much
shorter time scale $\tau_F$ is not resolved \cite{Alex_PRB2004}).
 For  low $T$ ($=0.1$ K) only states
with two possible energy values near the $E_F$ are populated. The
long-time behavior exhibits thus beatings and hence two peaks in
the emission spectrum (Fig.~\ref{Fig:echo_other} (b)) are
observable. With increasing $T$ more levels near  $E_F$ are
populated and the dynamics of the dipole moment  shows alternating
collapsed states and quantum revivals. The revival time is given
by $T_{\rm rev}=4\pi\hbar/|\partial^2 \varepsilon_m/\partial m^2|$
with $\varepsilon_m$ in the proximity of  $E_F$.
 The fast (classical) oscillations have a period $T_{\rm cl}=2\pi\hbar/|\partial
\varepsilon_m/\partial m|\approx \tau_{_F}$\cite{Robinett2004}.
The decay of the revival peak values is due to the relaxation. For
an estimate of the dipole-moment relaxation time, $\tau_{\rm
rel}$,
 we utilize  the decay dynamics and use the two first
consecutive envelope maxima to determine $\tau_{\rm rel}$ from a
fit by $A\exp(-t/\tau_{\rm rel})$. Fig.~\ref{Fig:echo_other}(c)
 displays the dependence of
$\tau_{\rm rel}$ on  $T$ and on the number of electrons $N$ in the ring.
$\tau_{\rm rel}$  increases rapidly with decreasing $T$ and, as expected,
$\tau_{\rm rel}$  tends to infinity for $T\to 0$. Fig.~\ref{Fig:echo_other} (b)
evidences that  by measuring the emission spectrum the temperature-dependent
decay dynamics can be traced. The different   decay dynamics of the dipole
moment and of the current (which can be monitored separately) allows insights
into different parts of the density matrix. The dipole moment (current) is
namely determined by the diagonal (near-diagonal) elements of the density
matrix.


 \emph{Magnetic field tuning.-}
Applying two time-delayed mutually perpendicular  HCPs generates
charge current $I$ and an associated magnetization $M$ in the ring
\cite{Alex_PRL2005}. The utilization of the pulse-induced magnetic
field to monitor locally, in a pump-probe manner, the ultra-fast
spin dynamics in nanostructures or to induce a fast ratchet effect
 require a fine control of the fast switching behavior of
the pulse-triggered magnetic field, a task which is tackled  here
for the first time by utilizing the collapsed states of the dipole
moments.
Two weak perpendicular pulses with strengths $\alpha_\parallel$
and $\alpha_\perp$ and delay time $\tau$  (chosen as $\tau=T_{\rm
cl}/4$) initiate along the respective field axis two independent
polarization dynamics $\mu_\parallel(t)$ and $\mu_\bot(t)$. Using
Eq.~\eqref{Eq:rho_basis_free_result} we derive that the charge
current change $\Delta I$ by application of the second HCP with
kick strength $\alpha_\bot$ at $t=t_{\rm sw}$ is given by
%
%
\begin{equation}\label{Eq:current}
   \Delta I=-\alpha_\bot\frac{\mu_\parallel(t=t_{\rm sw}^-)}{er_0}I_0,\: \ I_0=e\hbar/(m^*
   r_0^2).
\end{equation}
Correspondingly, the  ring magnetization is shifted  by $ \Delta M
= \pi r_{0}^{2}\Delta I$, and the current-generated magnetic flux
by $\Delta \Phi = \frac{\pi}{2}\mu_0 r_{0}\Delta I$ ($\mu_0$ is
the magnetic permeability of surrounding material).
  Eq.~\eqref{Eq:current} can be exploited  to
  turn the current on and off  producing thus short
  magnetic pulses, as follows: Having produced
  a current with a HCPs pair
   we subject (after the time $\tau_{p-p}$)
   the collapsed state to a second sequence of two pulses.
The first pulse of the second sequence has the  strength $\alpha'=\alpha_\parallel$ and
triggers the same dynamics of the dipole moment as does the first
HCP of the first HCPs pair because  the collapsed state is
unpolarized. The second pulse of the second sequence has strength $\alpha'_\bot$ and is
applied with polarity opposite to the second HCP of the first {pair}.
$|\alpha'_\bot|$ is chosen smaller than $\alpha_\bot$ by a factor determined by the
current decay during $\tau_{\rm p-p}$ and in such a way that the currents
triggered by the  {two HCP pairs} cancel  {each other}.
{Fig.~\ref{Fig:echo_other}(c)} demonstrates numerically the
accuracy of this scheme of current switching.
\begin{figure}[t]
  \centering
  \includegraphics[width=8.6cm]{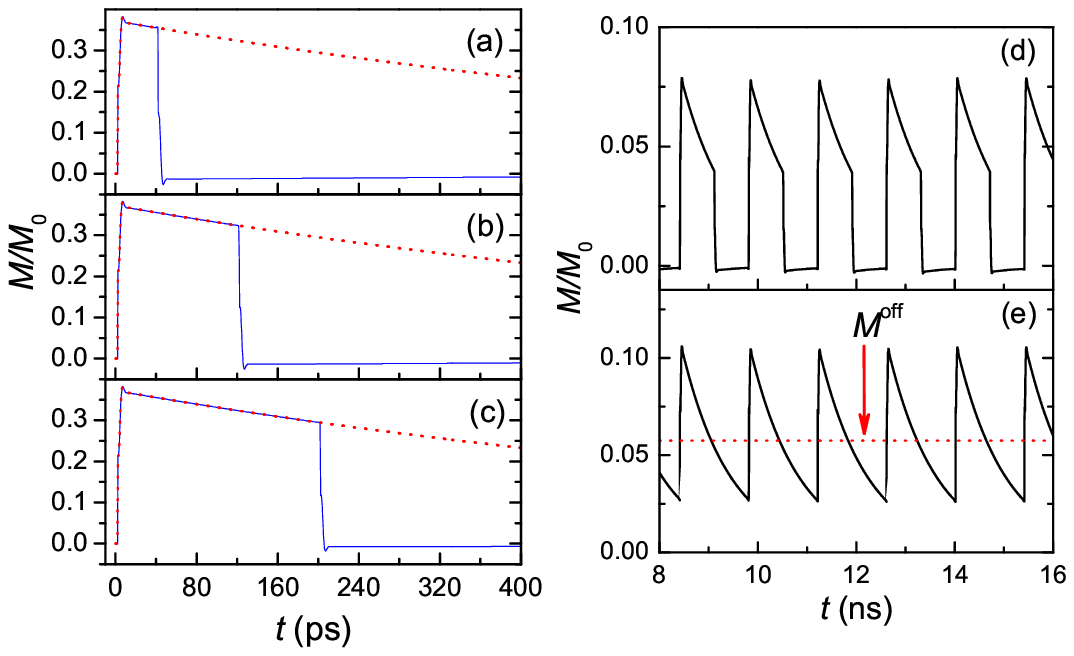}
  \caption{\label{Fig:current_control}
   {Magnetization pulses generated in the ring by application of two
  HCP pairs delayed by
  $\tau_{\rm p-p}=40$~ps in (a), $\tau_{\rm p-p}=120$~ps in (b), and $\tau_{\rm p-p}=200$~ps in (c).
  Kick strengths are $\alpha_\|=\alpha_\bot=\alpha'_\|=0.4$,
  $\alpha_\bot'=-0.39$ in (a), $\alpha_\bot'=-0.35$ in (b), and $\alpha_\bot'=-0.32$ in (c).
  Magnetization $M$ is normalized to $M_0=\pi r_0^2I_0$.
  Dotted lines show the result without applying the
  second HCP pair.
  Parameters of the calculation: $r_0=0.3~\mu{\rm m}$, $d=20~{\rm nm}$, N=160, T=20~K, $\tau_d=0.5$~ps.
  Figures (d) and (e) show periodically alternating time-asymmetric magnetization generated in the
  {ring} by
  application of series of sequences of HCPs with a time delay between sequences 1.4~ns.
  In (d) each sequence contains four HCPs having
  kick strengths $\alpha_\|=0.2,\alpha_\bot=0.2,\alpha'_\|=0.2,$ and $\alpha'_\bot=-0.1$, respectively.
  In (e) each sequence contains two HCPs having
  kick strengths $\alpha_\|=0.2$ and $\alpha_\bot=0.2$,
  respectively. $M^{\rm off}$ is the offset magnetization
  associated with an external magnetic field \cite{ratchet}.
  Parameters of the  {ring} and duration of HCPs as in Fig.~\ref{Fig:echo_other},
  $T=4$~K.}
   }
\end{figure}
The abrupt current switching opens the way for  the tunability of
the time-structure of the magnetic pulses. The  magnetic pulses
duration is stretchable depending on $\tau_{\rm p-p}$
(cf.~Fig.~\ref{Fig:current_control}(a)-(c)): As inferred from
Figs.~\ref{Fig:echo_other}(d),\ref{Fig:current_control}(a)-(c), if
 $\tau_{\rm p-p}\ll \tau_{\rm rel}^{\rm cur}$, where
$\tau_{\rm rel}^{\rm cur}$ is the time constant for the current
initial decay dynamics,  almost rectangular ps magnetic pulses are
produced. For $\tau_{\rm p-p} \gtrsim \tau_{\rm rel}^{\rm cur}$
asymmetric pulses are triggered, which is of importance when
studying ratchet effects \cite{ratchet}.
Fig.~\ref{Fig:current_control} demonstrates the level of
tunability of the pulse shape and duration by varying
experimentally accessible parameters. The magnitude of the
triggered pulses is enlarged for smaller rings and/or stronger
kicks. Also an appropriate arrangement of a collection of rings
allows additional tuning of the magnetic pulses.

In summary, we demonstrated how relaxation, collapses, and
 {revivals} in quantum rings can be studied by means of
electromagnetic pulses and how these phenomena can be exploited to
generate with current technology magnetic pulses with tunable time
and shape structure. Numerical calculations with realistic  pulse
and  material parameters are performed  endorsing the feasibility
of the predicted effects with nowadays technology.

\end{document}